\documentclass{article}

\usepackage{spconf} 
\usepackage{amsmath,graphicx}
\usepackage[english]{babel}
\usepackage[T1]{fontenc}
\usepackage{amsfonts, amssymb}
\usepackage{setspace}
\usepackage{pifont}
\usepackage[normalem]{ulem}
\usepackage{placeins}
\usepackage{xspace}
\usepackage[letterspace=10]{microtype}
\usepackage[colorlinks=true]{hyperref}
\usepackage{subcaption} % table figure
\usepackage{dsfont} % Indicator function and other math symbols
\usepackage{mathabx} % For \bigtimes
\usepackage{booktabs, multirow} % Tables
\usepackage{tablefootnote} % Footnotes in tables
\usepackage{xcolor} % Colors
\usepackage{dblfloatfix}    % To enable figures at the bottom of page

\title{Functional Invariants to Watermark Large Transformers}

\name{
Pierre Fernandez$^{1,2}$$^\star$,
Guillaume Couairon$^1$,
Teddy Furon$^2$$^\dag$,
Matthijs Douze$^1$
\thanks{
    \hspace{-1em}$^\star$Correpondance to \href{mailto:pfz@meta.com}{pfz@meta.com}
}    
\thanks{
    \hspace{-1em}$^\dag$\resizebox{0.97\linewidth}{!}{Work supported by ANR / AID under Chaire SAIDA ANR-20-CHIA-0011.}
}
}
\address{
    $^1$FAIR, Meta \qquad $^2$Centre Inria de l'Université de Rennes
}

\begin{document}

%% Comments
\newcommand{\rv}[1]{{\color{purple} [\textbf{Hervé}: #1]}}
\newcommand{\matthijs}[1]{{\color{orange} [\textbf{Matthijs}: #1]}}
\newcommand{\pf}[1]{{\color{blue} [\textbf{Pierre}: #1]}}
\newcommand{\teddy}[1]{{\color{red} [\textbf{Teddy}: #1]}}
\newcommand{\tofill}{{\color{orange} [\textbf{To fill}]}}

%% Math
\newcommand{\R}{\mathds{R}}
\newcommand{\N}{\mathds{N}}
\newcommand{\abs}[1]{\lvert{#1}\rvert}
\newcommand{\norm}[1]{\lVert{#1}\rVert}
\renewcommand{\vec}[1]{\mathbf{#1}}
\def\dotproduct#1#2{\langle#1, #2\rangle}
\def\1{\mathbf{1}}
\def\ind#1{\1_\mathrm{#1}}

\def\V{\mathcal{V}}
\def\A{\mathcal{A}}
\def\W{\mathcal{W}}
\def\G{\mathcal{G}}
\def\H{\mathcal{H}}
\def\Prob{\mathds{P}}
\def\E{\mathds{E}}
\def\key{k}
\def\temp{\theta}

%% Abrev
\newcommand{\etal}{\textit{et al}.\@ }
\newcommand{\eg}{\textit{e.g}.\@ }
\newcommand{\ie}{\textit{i.e}.\@ }
\newcommand{\aka}{{a.k.a}.\@ }
\newcommand{\muap}{$\mu$AP\@ }
\newcommand{\rone}{$R$@$1$\@ }
\newcommand{\Li}{ \mathcal L _{\mathrm{i}} }
\newcommand{\Lf}{ \mathcal L _{\mathrm{f}} }

%% Dings
\newcommand{\vardiamond}{\ding{169}}
\newcommand{\cmark}{\ding{52}}
\newcommand{\varcheck}{\ding{51}}
\newcommand{\xmark}{\ding{55}}
\newcommand{\varcross}{\ding{54}}
\newcommand{\acti}{\star}
\newcommand{\TODO}{{\color{blue}TODO}}

% Colors
\definecolor{apricot}{rgb}{0.98, 0.81, 0.69}

\ninept

\maketitle

\begin{abstract}
The rapid growth of transformer-based models increases the concerns about their integrity and ownership insurance. 
Watermarking addresses this issue by embedding a unique identifier into the model, while preserving its performance. 
However, most existing approaches require to optimize the weights to imprint the watermark signal, which is not suitable at scale due to the computational cost.
This paper explores watermarks with virtually no computational cost, applicable to a non-blind white-box setting (assuming access to both the original and watermarked networks). 
They generate functionally equivalent copies by leveraging the models' invariance, via operations like dimension permutations or scaling/unscaling. 
This enables to watermark models without any change in their outputs and remains stealthy.
Experiments demonstrate the effectiveness of the approach  and its robustness against various model transformations (fine-tuning, quantization, pruning), making it a practical solution to protect the integrity of large models.
\end{abstract}

\vspace{0.4em}
\noindent

\section{Introduction}

Large-scale transformer models are a leap forward in the field of machine learning, with large language models like GPT-4~\cite{openai2023gpt}, LLaMA~\cite{touvron2023llama} and others~\cite{mistral2023mistral, bard}, or vision ones like ViT-22b~\cite{dehghani2023scaling} or DINOv2~\cite{oquab2023dinov2}.
As these models grow in complexity and size, protecting them is important due to investments in their development.
Notably, this is raised by \href{https://www.whitehouse.gov/wp-content/uploads/2023/07/Ensuring-Safe-Secure-and-Trustworthy-AI.pdf}{\color{black}the US ``Ensuring Safe, Secure, and Trustworthy AI'' announcement}, \href{https://artificialintelligenceact.eu/}{\color{black}European AI Act} and \href{http://www.cac.gov.cn/2023-07/13/c_1690898327029107.htm}{\color{black}Chinese AI governance rules}.

Watermarking deep neural networks~\cite{uchida2017embedding, adi2018turning} presents a step towards ensuring their security, integrity and ownership. 
Embedding a unique identifier into the model enables tracing it to safeguard it from unauthorized usage and distribution.
However, watermarking large transformer models poses new challenges. 
Current watermarking methods involve optimizing the weights to infuse the watermark, either during pre-training or by fine-tuning the weights with additional losses.
While these techniques have shown success for smaller models, they become computationally infeasible for large-scale models and for the burgeoning number of potential users and applications.

To address these challenges, we introduce a new approach to watermarking large transformers, when access to both the original and watermarked model is granted, \ie in a non-blind white-box setting. 
Our method capitalizes on the inherent invariance of transformers. 
For a given model, it generates equivalent copies that serve as carriers for arbitrary signatures.
By employing operations such as dimension permutation and coupled matrix multiplications, we create model replicas without changing the model's outputs and without training.
We conduct experiments on state-of-the-art transformer architectures to evaluate the applicability of our approach and its robustness against model processing (\eg fine-tuning, pruning, quantization, etc.).
We also discuss the main drawbacks of this setting. 

\begin{figure}[t]
    \centering
    \vspace{0.2em}
    \includegraphics[width=1.0\linewidth]{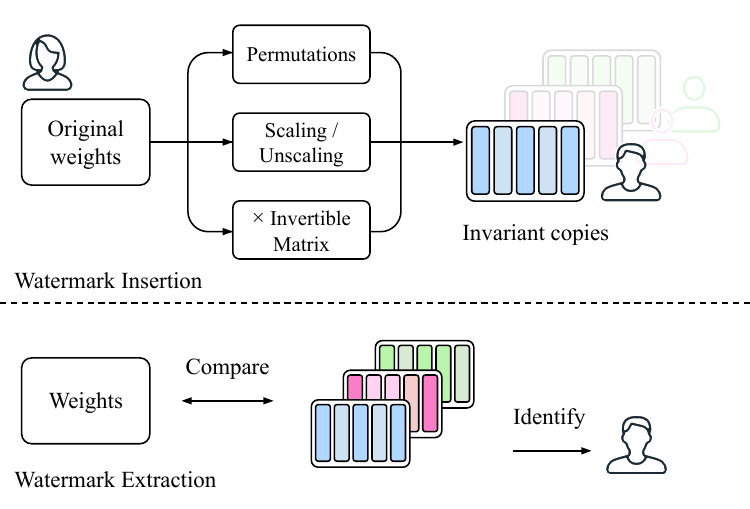}
    \caption{
        Overview. 
        We identify each model by applying invariance operations to the original weights.
    }
    \label{fig:overview}
\end{figure}

The paper is organized as follows: 
\autoref{sec:related} provides an overview of related works on DNN watermarking and background on transformers;
\autoref{sec:method} details the transformer's invariants and how to exploit them for watermarking;
\autoref{sec:experiments} presents experimental results on large language models.

\vspace*{-1em}
\paragraph*{Problem statement.}
A provider \emph{Alice}, distributes her model to various users \emph{Bob} (either individuals or organizations).
She aims to trace the model back to a specific user, in case of unauthorized distribution or leaks. 
As a precautionary measure, Alice embeds a unique signature in the model's weights for each user. 
In a white-box setting, Alice has access to the models' weights and extracts the signature from it to identify Bob.
Besides, Bob may evade detection intentionally (trying to remove the watermark) or unintentionally (fine-tuning, quantization, etc.).

This setting is quite common. 
Indeed few entities (``Alices'') have the necessary computation resources, data and expertise to generate the base model. 
For example, the training of the 65-B LLaMA model took around $1$B GPU-hours.
Therefore, there are few variants of such large models in the world. 
Besides, when Bob gains access to the base model, it is common that he transforms it and that it re-emerges in a public forum or through another channel, so that Alice can analyze it. 
This can be either because Bob re-distributed it or because Alice sought the model through legal channels, as suggested by Fan \etal~\cite{fan2021deepip}.
For example, many variants of the LLaMA models have been fine-tuned on instruction datasets and been made available online.

\newcommand{\att}[1]{\mathrm{Att}^{#1}}
\newcommand{\ffn}[1]{\mathrm{Ffn}^{#1}}
\newcommand{\lnatt}[1]{\mathrm{Ln}_{\mathrm{att}}^{#1}}
\newcommand{\lnffn}[1]{\mathrm{Ln}_{\mathrm{ffn}}^{#1}}
\newcommand{\lnout}{\mathrm{Ln}_{\mathrm{out}}}

\section{Related Work \& Technical Background}\label{sec:related}

\subsection{Deep Neural Network (DNN) Watermarking}\label{par:watermarking} 

DNN watermarking robustly embeds a unique identifier into the model without affecting its performance, in order to later verify the model's identity.
Watermarking should satisfy three criteria, 
\textit{utility}: the new model should have the same performance as the original one;
\textit{security}: it should be stealthy, hard to remove and to forge;
\textit{robustness}: the watermark should be detected even after the model has been modified.
Modifications may be unintentional -- models are fine-tuned, pruned and quantized
-- or intentional -- adversaries may try to remove the watermark or embed their own~\cite{fan2019rethinking, zhang2020passport, kallas2022rose}.
For instance, some adversarial transforms employ invariance operations in neurons and ReLU layers to evade detection~\cite{yan2023rethinking}, in a similar fashion as the techniques of this work.

We distinguish between white-box and black-box settings, depending on whether the model weights are accessible at verification time, or only through a remote API.
In white-box, the pioneering work~\cite{uchida2017embedding} embeds watermarks into the DNN's weights. 
A regularization loss term during training constrains the weights to carry a specific signal, while minimizing the impact on the model's performance. 
The watermark is then retrieved directly by analyzing the model's weights.
The  Deep·Signs·Marks~\cite{darvish2019deepsigns,chen2019deepmarks} extends this to target black-box settings and propose building collusion-resistant watermarks, 
RIGA~\cite{wang2021riga} improves its covertness and robustness, 
and greedy residuals~\cite{liu2021watermarking} improves the selection of the weights to modify.

Another line of work, called trigger-set based methods, embeds the watermark in the behavior of the model with regards to certain inputs. 
A recurrent idea is to use ``backdoors'', \ie memorize certain sequences of input-output pairs~\cite{adi2018turning, zhang2018protecting}. 
Watermarking generative models is also an active field of research, either by employing triggers~\cite{lim2022protect, ong2021protecting}, or by watermarking their outputs~\cite{fernandez2023stable, kim2023wouaf}.

The literature on watermarking large models is scarce, and none of the current papers operate at our scale. 
The most recent works~\cite{liu2023trapdoor, jiang2023ipcert} concentrate on ResNet-18/AlexNet ($\approx$20M parameters). 
PLMmark~\cite{li2023plmmark} also needs training and evaluates at most on BERT-large which is around 300M parameters. 
This is 100 times smaller than the models we consider in our work (\eg LLaMA-30B, LLaMA-70B). 
Previous methods could be adapted in the context of LLMs, but all of them would require training or at least fine-tuning. 
Their impact on the quality of the text generation and the robustness of the watermark is also not demonstrated. 
Thus, the feasibility of existing watermarking methods to these models remains an open question.

\vspace{-1em}
\subsection{Transformers}

Transformer~\cite{vaswani2017attention} neural networks have become the standard for many applications in the last few years.
They can be trained efficiently on GPUs and scale well to large datasets and models, in both natural language processing~\cite{raffel2020exploring, kaplan2020scaling} and computer vision~\cite{dehghani2023scaling, oquab2023dinov2}.
In the following we describe the NLP architecture from~\cite{radford2019language}.

The input string is first tokenized into a sequence of integers $(x_1, \dots, x_n) \in \V^n$.
An embedding layer $E \in \R^{|\V|\times d}$ maps each token $x_i$ to a continuous vector $z_i^0 = E_{x_i} \in \R^d $, where $d$ is the embedding dimension.
The transformer is a stack of attention and feed-forward layers, that we describe in the following.

\vspace{-1em}
\paragraph*{Attention layers.} 
The self-attention mechanism enables long-range dependencies between sequence elements.
A self-attention transforms an input sequence $\vec{z} \in \R^{n \times d}$ into queries $Q$, keys $K$, and values $V$:
\begin{equation}
    \resizebox{0.91\linewidth}{!}{
        \hspace*{-0.05\linewidth}
        $Q = \vec{z}W^\mathrm{Q} \in \R^{n \times d_\mathrm{k}};\;
        K = \vec{z}W^\mathrm{K} \in \R^{n \times d_\mathrm{k}};\;
        V = \vec{z}W^\mathrm{V} \in \R^{n \times d_\mathrm{v}}$.
    }
\end{equation}
It then computes attention weights by taking a scaled dot product between the queries and keys:
\vspace{-0.3cm}
\begin{align}\label{eq:attention}
    \mathrm{Attention}(Q,K,V) = \mathrm{softmax} \left( \frac{QK^\top}{\sqrt{d_\mathrm{k}}} \right)V.
\vspace{-0.3cm}
\end{align}
Where the $\mathrm{Softmax}$ operator is applied column-wise.

This attention operator is applied $h$ times in parallel, yielding $h$ output \emph{heads}.
The results are concatenated and projected back to the original dimension:
\vspace{-0.1cm}
\begin{align}\label{eq:multihead}
\mathrm{MultiHead}(Q,K,V) = \mathrm{Concat}(\mathrm{head}_1,., \mathrm{head}_h)W^\mathrm{O},
\vspace{-0.3cm}
\end{align}
where $\mathrm{head}_i = \mathrm{Attention}(QW_i^\mathrm{Q}, KW_i^\mathrm{K}, VW_i^\mathrm{V})$. 
The projections $W_i^\mathrm{Q}, W_i^\mathrm{K} \in \R^{d \times d_\mathrm{k}}$, $W_i^\mathrm{V} \in \R^{d \times d_\mathrm{v}}$ and $W^\mathrm{O} \in \R^{hd_\mathrm{v} \times d}$ are learned.

\vspace{-0.3cm}
\paragraph*{The feed-forward network.}
The output is fed to a feed-forward network (FFN), \eg two linear layers with a ReLU activation:
\vspace{-0.1cm}
\begin{align}
\mathrm{FFN}(\vec{h}) = \mathrm{ReLU}(\vec{h}W_1 + b_1)W_2 + b_2,
\vspace{-0.3cm}
\end{align}
where $W_1 \in \R^{d \times d_\mathrm{ff}}$, $b_1 \in \R^{d_\mathrm{ff}}$, $W_2 \in \R^{d_\mathrm{ff} \times d}$ and $b_2 \in \R^{d}$ are learned parameters (SwiGLU~\cite{shazeer2020glu} and other variants also frequently replace ReLU).

\vspace{-0.3cm}
\paragraph*{A stack of residual connections.}
Instead of directly feeding $\vec{z}$ and $\vec{h}$ to the attention and FFN layers, residual connections are applied and the inputs are normalized using layer normalization~\cite{ba2016layer} (or variants like RMSnorm~\cite{zhang2019root}):
$\label{eq:layernorm}
\mathrm{LayerNorm}(\vec{z}) = \frac{\vec{z} - \mu}{\sigma} \odot g + b,
$
where $\mu$ and $\sigma$ are the mean and standard deviation of $\vec{z}$ along its second dimension, and $g\in \R^d$ and $b\in \R^d$ are learned parameters.
This is repeated for each layer $l\in \{1, ..., L\}$ of the transformer:
\vspace{-0.1cm}
\begin{align}
\vec{h}^{l} &= \att{l} \left( \lnatt{l} \big( \vec{z}^{l} \big) \right) + \vec{z}^{l} \\
\vec{z}^{l+1} &= \ffn{l} \left( \lnffn{l} \big( \vec{h}^{l} \big) \right) + \vec{h}^{l}.
\vspace{-0.3cm}
\end{align}
The output is fed to a normalization layer $\lnout$ and a linear layer $W_\mathrm{out} \in \R^{d \times |\mathcal V|}$ to generate logits, and a softmax outputs the probability distribution of the next token.

\vspace{-0.3cm}
\paragraph*{Positional embeddings.} 
For many tasks, it is useful to encode the position of tokens in the input sequence.
Positional embeddings are what allows to encode this information.
They were originally sinusoidal functions of the position~\cite{vaswani2017attention} added to the input embeddings. 
There are now several variants~\cite{raffel2020exploring, su2021roformer, press2021train, kazemnejad2023impact}, that may change Eq.~\eqref{eq:attention}.
For instance, rotary embeddings~\cite{su2021roformer} multiply queries and keys depending on their relative position in the sequence.
If $m$ is the position of the query ($Q_m= \vec{z}_m W^\mathrm{Q}$) and $n$ the position of the key, then it rewrites the product of~\eqref{eq:attention} as:
\vspace{-0.1cm}
\begin{align}\label{eq:rotary}
    \qquad Q_m K^\top_n = z_m W^\mathrm{Q} R_{\Theta, n-m} (z_n W^\mathrm{K})^\top.
\vspace{-0.3cm}
\end{align}
$R_{\Theta,n}$ is a block diagonal matrix with $2\times2$ rotation matrix entries:
\vspace{-0.1cm}
\begin{align*}
   \left( R_{\Theta,n} \right) _i  =
   \begin{pmatrix}
   \cos n \theta_i & -\sin n \theta_i \\
   \sin n \theta_i & \cos n \theta_i
   \end{pmatrix},
\vspace{-0.3cm}
\end{align*}
with rotation frequencies chosen as $\theta_i = 10,000^{-2i/d}$.

\begin{figure*}[b!]
    \centering
    \includegraphics[width=1.0\linewidth]{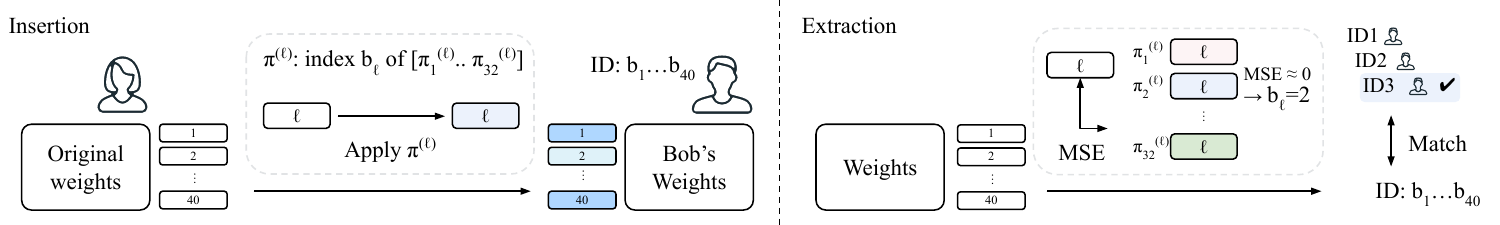}
    \caption{
    Detailed illustration of watermark insertion and extraction, with the example of permutation on $L$=40 blocks. 
    A user ID is a list $b_1...b_L$ of $L$ bytes, that are used to select the permutation to apply for each block $\ell$.
    For each $\ell$, the extraction computes the MSE between the observed weights and all original permuted weights. 
    It then selects the one with minimum MSE, which in turn gives $b_\ell$.
    }
    \label{fig:method}
\end{figure*}

\section{Watermarking through Invariance}\label{sec:method}

\subsection{Invariants in the weights of transformers}
We define an invariant as a series of operation applied on the model's weights $\theta \rightarrow \theta ' $ such that for any input $x$, the output $f_{\theta'}(x)$ is the same as before the application of the invariant.

\vspace*{-1em}
\paragraph*{Permutation invariance} appears in (at least) four levels of the transformer.
We note $\Pi^d$ the set of permutations of $\{1,..., d\}$. 
For a matrix $M\in\R^{d\times d}$ and $\pi \in \Pi^d$, we denote by $M_{:,\pi}$ (resp. $M_{\pi,:}$) the matrix where columns (resp. rows) are permuted according to $\pi$.

\vspace*{0.2em}
\noindent
\emph{Embedding dimension.} 
The embedding matrix $E$ can be permuted along its second dimension without changing the output of the model, as long as the permutation is propagated to other matrices in the model. 
More formally, for $\pi \in \Pi^d$, if $E' = E_{:,\pi}$, then matrices $\{ W^\mathrm{Q}, W^\mathrm{K}, W^\mathrm{V}, W_1, W_{out}, \lnatt{}, \lnffn{}, b_2 \} \subset \theta$ need to be permuted along their first dimension by $\pi$ and all matrices $\{ W^\mathrm{O}, W_2 \}$ along their second one: 
$(W^\mathrm{Q})' = W^\mathrm{Q}_{\pi, :}$, $(W^\mathrm{O})' = W^\mathrm{O}_{:, \pi}$, etc.

\vspace*{0.2em}
\noindent
\emph{FFN layer dimension.}
All neurons making up matrices $W_1$ and $W_2$ of feed-forward networks can be permuted:
for $\pi \in \Pi^{d_\mathrm{ff}}$, if $W_1' = (W_1)_{:,\pi}$ and $W_2' = (W_2)_{\pi,:}$, then $f_{\theta'}(\cdot) = f_{\theta}(\cdot)$.

\vspace*{0.2em}
\noindent
\emph{Attention heads.}
Heads are interchangeable in~\eqref{eq:multihead} provided that $W^\mathrm{O}$ is permuted in blocks of $d_\mathrm{v}$ according to its first dimension.
% In practice, the $(W_i^\mathrm{Q})_{i \in \{1,..,h\} }$ are stored as a unique matrix, so permuting the heads is equivalent to permuting the columns of this matrix in blocks of $d_\mathrm{k}$ (same for $W_i^\mathrm{K}$ and $W_i^\mathrm{V}$).

\vspace*{0.2em}
\noindent
\emph{Inside the head.}
Depending on the type of positional embeddings, the previous permutations can be extended.
For instance if they do not impact~\eqref{eq:attention} (this is not the case for rotary embeddings) then $W^\mathrm{Q}$ and $W^\mathrm{K}$ can be permuted along their second dimension.

\vspace{-0.3cm}
\paragraph*{Scaling/Unscaling.}\label{sec:scaling}
Whenever layer norms or variants are directly followed (or preceded) by linear layers, \eg at every attention or FFN block, we can rescale component-wise the parameters $g$, $b$ of $\mathrm{LayerNorm}(\vec{z})$
%Eq.~\ref{eq:layernorm}
by a vector $\alpha\in \R^d$.
Invariance is obtained by dividing the rows of the following (or preceding) linear layers by the same vector.

\vspace{-0.3cm}
\paragraph*{Invertible matrices in QK products.}\label{sec:invertible}
We hereby assume the positional embeddings do not impact~\eqref{eq:attention}.
If $P \in \R ^{d_\mathrm{k}\times d_\mathrm{k}}$ is invertible, then choosing $(W^\mathrm{Q})' = W^\mathrm{Q} P$ and $(W^\mathrm{K})' = W^\mathrm{K} (P^\top)^{-1}$ is invariant in~\eqref{eq:attention}.
This also applies to the case of rotary embeddings by restricting $P$ to be block diagonal of $2\times 2$ matrices that apply a 2D rotations and scaling by a factor $\lambda$ (thanks to the commutativity of 2D rotations).

\vspace{-0.3cm}
\paragraph*{Combining invariants.}
All previous parameter transformations may be seen as invertible right or left matrix multiplications applied to the model parameters. 
% For instance, $\pi$ is associated to a permutation matrix $P$ such that $P_{i, \pi(i)} = 1$ and $P_{i, j} = 0$ otherwise, scalings are diagonal matrix, etc.
They do not interfere and may be combined in a sequence of arbitrary order, yielding $\theta \rightarrow \theta' \rightarrow \theta'' \rightarrow \cdots$.

Combining transformations at all levels improves robustness to intentional removal attacks and to collusion (\ie when several Bobs share their weights to evade detection).
Indeed, if Bob tries to remove the watermark by re-applying one invariant, it will still be present in the other invariants.
In the same way, if several Bobs compare their models, it will be hard for them to identify which operations were applied to their models, since the order in which they were applied is unknown, and since the weights will differ a lot between them.

\subsection{From invariants to watermarks}\label{sec:invariantsaswatermarks}

\paragraph*{Insertion.}
Before starting the watermark process, for each invariant and each level of the network, we restrict the set of transformations to $2^k$.
For example, we randomly sample $2^k$ possible permutations in $\Pi^d$ for the Embedding dimension (out of the total $d!$). 
% This is to simplify the detection procedure. 

Therefore, we can encode $k$ bits for each combination of an invariant and a level. 
We encode a model's identifier as the concatenation of $m$ chunks of $k$ bits ($2^{mk}$ possibilities).
For instance, let $k=4$ and the model have $32$ layers. 
We choose to embed two permutations per layer, one for the attention block and one for the FFN block. 
The total number of bits is $2\times 32\times 4=256$, representing $10^{77}$ possible models (approximately the number of atoms in the universe, an upper bound of the number of Bobs). 
% This embedding process provides robustness and efficiency when extracting the watermarks.

\vspace*{-1em}
\paragraph*{Extraction.}
To extract the $k$-bits message from a weight matrix, we re-apply all $2^k$ possible invariants to the original matrix.
We then compute the Frobenius norm of the difference (MSE) between the observed weight and the possible watermarked ones.
We choose the one with lowest MSE among the $2^k$ and this choice is encoded as a $k$-bit integer.
Doing that on every blocks of the network and every invariant, we end up with a full message made of $m$ chunks of $k$ bits.
In the case of intertwined invariants, we do the same in the order in which we inserted the invariants, reverting them as we go.

To speed up extraction, we may select a subset of the matrices' rows before extraction.
This speeds up the extraction (in the order of 100$\times$), but makes the detection slightly less robust. 
For instance, in the case of scaling/unscaling we may select the first $100$ components of $\alpha$ from $\R^{d}$ to $\R^{100}$ and $W$ from $\R^{d\times d'}$ to $\R^{100\times 100}$.  

\vspace*{-1em}
\paragraph*{Matching.}
To match an extracted message (made of $m$ chunks of $k$ bits) with a model's identifier, we compute the number $s$ of chunk-wise errors with respect to all possible identifiers.
We return a match if $s$ is bellow a fixed threshold $\tau$ to ensure resilience to corruption and to provide a confidence score.
A theoretical p-value, \ie the probability of obtaining a number of errors lower than $s$ for a random model, is given by the regularized incomplete beta function $\mathcal{I}$:
\vspace{-0.5em}\begin{equation}\label{eq:pvalue}
    \textrm{p-value}(s) = 1- \left(1-\mathcal{I}_{ 1/2^{k} } ( m-s, s+1) \right)^N,
\vspace{-0.1cm}
\end{equation}
where $N$ is the number of distributed models. 
% $N$ accounts for the fact that when multiple models are distributed, the risk of finding a match by chance is higher.
% (birthday paradox).

\vspace*{-1em}
\paragraph*{Robustness and security.}
Watermarking models through invariance is stealthy, because it does not change their outputs.
However, a distortion-free watermark is also a weakness: Alice can hide the watermark without impacting the model's utility, but on the other hand an adversarial Bob may do the same at no cost.
In short, most of these watermarks are very robust against classical model manipulations (fine-tuning, quantization, etc.) but not against a malicious user who knows the method.
In this case we would only know that the model is an unauthorized copy, without knowing the leaker.

\newcommand{\valtab}[1]{{\color{gray}\footnotesize ({#1})}}

\section{Experiments}\label{sec:experiments}

The purpose of the experiments is to evaluate the effectiveness and the robustness of the watermarks to transformations on transformers, in the context of large language models.

\vspace*{-1em}
\subsection{Setup}

\paragraph*{Model.}
We use LLaMA~\cite{touvron2023llama} models as main benchmark. 
The architectural differences with regards to the original transformer architecture are pre-normalization~\cite{radford2019language} with RMSnorm~\cite{zhang2019root}, SwiGLU activation~\cite{shazeer2020glu} and rotary embeddings~\cite{su2021roformer}.
To evaluate that the utility of the model is not degraded, we show results on a next-token prediction task.
This is done on random sequences of text taken from Wikipedia, then tokenized using the default tokenizer of LLaMA.
Unless stated otherwise, we use the $7$B-parameter model.

\vspace*{-1em}
\paragraph*{Attacks.}
We consider the following attacks.
\emph{Fine-tuning.} 
We fine-tune the model in a supervised manner with the same settings as~\cite{alpaca}, on 3 epochs with learning-rate of $2\times 10^{-5}$.
\emph{Noise.} 
We add zero-mean Gaussian noise with standard deviation $\sigma$ to the model weights.
\emph{Quantization.}
We quantize the model weights into $b$ bits. 
To allow flexible rates and ease the experiments, this is done by uniformly quantizing the weights between their minimum and maximum values.
\emph{Pruning.}
We prune the model weights by zeroing the ones with smallest L1 norms, with sparsity given in percentage of zero weights.

\vspace*{-1em}
\paragraph*{Watermark settings.}
We apply the encoding process of Sect.~\ref{sec:invariantsaswatermarks}.
For permutation invariance, we permute attention heads and FFN layers.
For scaling, we alter the layers' RMSnorms and following matrices.
The scaling vector $\alpha$ is such that $\log_{10}(\alpha) \sim \mathcal{U}(-1, 1)$.
For QK products, as mentioned in Sect.~\ref{sec:invertible}, the invertible matrix has to be block diagonal of $2$ by $2$ rotation matrices, so we randomly sample $d/2$ rotation angles.
We fix the number of possible choices at $k$=$8$, \ie each choice is encoded with a byte.
Therefore, we encode $2$ bytes at every layer, except in QK products where we encode $1$.
When combining all invariants together, we proceed the same way for all blocks: we start with permutation, then apply invertible matrices in QK products, then scale/unscale the layer norms and matrices.

For instance, the $7$B model has $L$=$32$ layers so the watermark is $64$ bytes long except for the QK products invariant where it is $32$.
In the case of combined invariants, the total number of bytes is $160$.

\subsection{Results.}\label{sec:results}

\begin{table}
    \centering
    \caption{
    Distortion induced on generation 
    % (measured as the proportion of generated tokens that differ between the original and the watermarked model);
    and robustness of watermark extraction under various processes. 
    Each line stands for a different invariant.
    We present results of the sped-up extraction, the ones for no speed-up are given as \valtab{acc}.
    \vspace*{-0.5em}
    }
    \label{tab:robustness}
    \resizebox{1.0\linewidth}{!}{
        \begin{tabular}{lc *{4}{l@{\hspace{4pt}}}  }
            \toprule
            \multirow{2}{*}{Method} & \multirow{2}{*}{Distortion} & \multicolumn{4}{c}{Byte accuracy ($\%$) on:} \\
            \cmidrule(lr){3-6}
                        &           & Noise $1.0$    & Quant. $3$b & Prun. $50\%$ & Fine-tune  \\
            \midrule
            Perm. & 0.20$\%$ & 51.4  \valtab{99.6} & 72.0 \valtab{100.0}   & 100.0     & 100.0 \\
            QK          & 0.18$\%$ & 100.0          & 100.0         & 100.0     & 100.0 \\
            Scaling     & 0.24$\%$ & 100.0          & 98.1 \valtab{100.0}       & 100.0     & 100.0 \\
            All        & 1.77$\%$ & 60.8  \valtab{99.8} & 70.0   \valtab{99.4} & 100.0 & 100.0 \\
            \bottomrule
        \end{tabular}
    }
\end{table}

\paragraph*{Robustness.}
We evaluate the robustness of the watermark using the \emph{byte accuracy}, \ie the percentage of bytes correctly recovered.
Results are averaged over $N$=$100$ watermarked models except for fine-tuning where we only fine-tune one model.
We speed-up the extraction by selecting a subset of $100$ rows of the matrices (see Sect.~\ref{sec:invariantsaswatermarks}); 
time needed for extraction is around 20 minutes when using the full matrix instead.

Table~\ref{tab:robustness} reports the byte accuracy for different processing applied before extraction.
We observe that the watermark is robust to all attacks with byte accuracy >$50\%$.
Errors mainly come from the speed-up of the extraction process.
We also consider the \emph{p-value} of the associated statistical test~\eqref{eq:pvalue}.
A byte accuracy of 50\% on 64-bytes messages is more than enough to reliably identify a model: 
the p-values are always bellow $10^{-60}$, due to the very low probability of simultaneously observing a match between tens of pairs of random bytes.
As an illustration, 8 matching bytes on 64-bytes messages already gives a p-value of $10^{-8}$.

\vspace*{-1em}
\paragraph*{Model's utility.}
In fact, previous invariants are not perfect because of quantization (weights are stored as 16bits floating point numbers).
Thus, we quantitatively compare watermarked and original models.
We feed to both of them $1$k sequences of $256$ tokens.
Predicted next tokens are greedily chosen as the argmax of the $256$k observed logits.

Table~\ref{tab:robustness} reports the distortion as the proportion of predicted tokens that differ between watermarked and original models.
As expected this proportion is very low (<$1.8\%$) and higher for the scaling invariant since it further affects quantization.
Besides, the distortion increases when the token is far in the sequence \eg for sequences of length 1024 tokens, the average distortion at the last token rises to $2.5\%$ for the scaling invariant.
This is still very low and does not affect the utility of the model since predicted tokens are still likely.

\vspace*{-1em}
\paragraph*{Computational efficiency.}
Larger models have more layers and parameters, which increases the computational cost of insertion and extraction.
In Table~\ref{tab:modelsize}, we report results for different model sizes.
Insertion and extraction times are averaged over $100$ runs and measured on 2 Intel(R) Xeon(R) 6230 @ 2.10GHz cores and a total of 480GB of RAM.
The low computational costs and requirements (no GPU needed) makes it possible to scale to very large models.

\begin{table}
    \centering
    \caption{
        Computational cost of watermark insertion and extraction for different model sizes and the different invariants. 
        \vspace*{-0.5em}
    }
    \label{tab:modelsize}
    \resizebox{\linewidth}{!}{
        \begin{tabular}{r@{\hspace{8pt}}c@{\hspace{8pt}}c  *{3}{r@{\hspace{4pt}}} @{\hspace{8pt}}  *{3}{r@{\hspace{4pt}}}}
            \toprule
            \multirow{2}{*}{Model} & \multirow{2}{*}{$L$} & \multirow{2}{*}{$d$} &  \multicolumn{3}{c}{\hspace{-12pt}Insertion (s)} & \multicolumn{3}{c}{Extraction (s)} \\
            & & & Perm. & Scaling & QK   & Perm. & Scaling & QK  \\
             \cmidrule(rr{12pt}){4-6} \cmidrule(rr){7-9}
            7b & 32 &  4096  & 3.5  & 2.7 &    7.4    &  9.2 &  31.7 &  6.0    \\
            13b & 40 & 5120  & 7.0  & 4.9 &    15.8   & 14.1 &  30.3 &  7.7    \\
            30b & 60 & 6656  & 19.3 & 8.7 &    47.3   & 31.7 &  54.7 & 13.5    \\
            70b & 80 & 8192  & 37.1 & 17.5 &   106.0 &   56.3 & 110.0 & 21.5   \\
            \bottomrule
        \end{tabular}
    }
    \vspace*{-0.5em}
\end{table}

\section{Conclusion}

Our work presents a lightweight approach for watermarking large transformers. 
We leverage invariance properties to generate equivalent copies for watermark embedding. 
It ensures that the model's outputs are preserved while providing close-to-perfect robustness against processes like fine-tuning or quantization. 

Yet, this approach has limitations. 
Namely, it is limited to white-box scenarios. 
Additionally, if a sophisticated attacker identifies all invariants, they may remove the watermark by applying the same transformation techniques. 
In this case, it would still be possible to identify that the model is an unauthorized copy but without the corresponding binary signature.
Overall, this work is a starting point to exploit invariance properties that stem from the extreme redundancy of parameters of large networks, for watermarking applications.

\bibliographystyle{ieeetr}
\begingroup
    \footnotesize
    \begin{spacing}{0.9} % Reduce line spacing
        \lsstyle % see microtype, allows to reduce character spacing
        \bibliography{references}
    \end{spacing}
\endgroup

\end{document}